\documentclass[aps,prd,twocolumn,superscriptaddress,amsmath,amssymb,showpacs]{revtex4}
\usepackage{color}
\usepackage{graphics}

\begin{document}

\title{Comment on ``Torsion Cosmology and the Accelerating Universe''}%

\author{A.~V.~Minkevich}

\email{minkav@bsu.by, awm@matman.uwm.edu.pl}

\affiliation{Department of Theoretical Physics, Belarussian State
University, Minsk, Belarus}

\affiliation{Department of Physics and Computer Methods, Warmia
and Mazury University in Olsztyn, Poland}

\author{A.S. Garkun}

\email{garkun@bsu.by}

\affiliation{Department of Theoretical Physics, Belarussian State
University, Minsk, Belarus}

\author{V.I. Kudin}

\affiliation{Department of Technical Physics, Belarussian National
Technic University, Minsk, Belarus}

\date{\today}

\begin{abstract}
Cosmological solutions for homogeneous isotropic models in the framework of the Poincar\'e gauge theory of gravity based on gravitational Lagrangian adopted in the paper by Kun-Feng Shie, James M. Nester and Hwei-Jang Yo (Phys. Rev. D \textbf{78}, 023522 (2008)) are discussed. Cosmological solutions for accelerating Universe obtained in referred paper are compared with corresponding solutions of standard $\Lambda CDM$-model and also with cosmological solution obtained by authors of this Comment. 
\end{abstract}


\maketitle
The referred paper \cite{1} is devoted to investigation of
isotropic cosmology in the frame of the Poincar\'e gauge theory of
gravity (PGTG) based on gravitational Lagrangian, which contains
besides scalar curvature also the term quadratic in the scalar
curvature and three terms quadratic in the torsion tensor with
some coefficients. The PGTG with such gravitational Lagrangian was
adopted in previous works of authors as describing dynamical
torsion field. According to \cite{1} corresponding theory can give
the solution of the problem of the acceleration of cosmological
expansion at present epoch (dark energy problem of general
relativity theory (GR)). Note that this problem was investigated
in \cite{2} in the frame of PGTG by using general expression for
gravitational Lagrangian including both a scalar curvature and
various terms quadratic in the curvature and torsion tensors with
indefinite coefficients, where regular inflationary Big Bang
scenario with accelerating stage of cosmological expansion at
asymptotics was proposed. Because the behaviour of cosmological
solutions for accelerating Universe obtained in \cite{1} and
\cite{2} are essentially different, we will analyze below in what
relation to cosmological solution obtained in \cite{2} is the
cosmological solution of \cite{1}. With this purpose we will
consider the application of gravitational equations of PGTG for
homogeneous isotropic models (HIM) deduced in \cite{2} in the case
of gravitational Lagrangian used in \cite{1}.

The following expression for gravitational Lagrangian of PGTG was
used in \cite{2} (by applying notations and definitions of
\cite{2}) \footnote [11] {In order to compare equations in
\cite{1,2} relations for indefinite parameters of gravitational
Lagrangians used in \cite{1,2} are given below (parameters of
\cite{1} are denoted by means of the prime): $f_0=-\frac{a_0'}
{{2}}$, $f_6=\frac{b'} {{24}}$, $a_1= - \frac{a_1'} {{2}}$,
$a_2=-a_1'$, $a_3=2a_1'$, coefficients $f_i$ $(i=1,2,...5)$ are
equal to zero in \cite{1}.}:
\begin{eqnarray}\label{1}
\mathcal{L}_{\rm g}=  f_0\,
F+F^{\alpha\beta\mu\nu}\left(f_1\:F_{\alpha\beta\mu\nu}+f_2\:
F_{\alpha\mu\beta\nu}+f_3\:F_{\mu\nu\alpha\beta}\right)  \nonumber \\
+ F^{\mu\nu}\left(f_4\:F_{\mu\nu}+f_5\: F_{\nu\mu}\right) +
f_6\:F^2 \nonumber \\
+S^{\alpha\mu\nu}\left(a_1\:S_{\alpha\mu\nu}+a_2\:
S_{\nu\mu\alpha}\right)
+a_3\:S^\alpha{}_{\mu\alpha}S_\beta{}^{\mu\beta}. 
\end{eqnarray}
In general case any HIM in PGTG is characterized by three
functions of time: the scale factor of Robertson-Walker metrics
$R$ and two torsion functions $S_{1}$ and $S_{2}$. Then not
vanishing components of the curvature tensor can be expressed by
the following 4 functions:
\begin{eqnarray}\label{2}
    A_1=\dot{H}+H^2-2HS_1-2\dot{S}_1,
    \nonumber\\
    A_{2}  = \frac{k} {{R^2 }} + \left( {H - 2S_1 } \right)^2  - S_2^2,
    \nonumber\\
    A_{3}  = 2\left( {H - 2S_1 } \right)S_2,
    \nonumber\\
    A_{4}  = \dot S_2+HS_2,
\end{eqnarray}
where $H=\dot{R}/R $ is the Hubble parameter and a dot denotes the
differentiation with respect to time. The total system of
gravitational equations of PGTG for HIM filled with spinless
matter with energy density $\rho$ and pressure $p$ takes the
following form \cite{2}:
\begin{eqnarray}\label{3}
a\left( {H - S_1 } \right)S_1  - 2bS_2^2  - 2f_0 A_{2}  + 4f\left(
{A_{1}^2 - A_{2}^2 } \right) \nonumber \\ + 2q_2 \left( {A_{3}^2
- A_{4}^2 } \right) =  - \frac{\rho } {3},
\end{eqnarray}
\begin{eqnarray}\label{4}
a\left( {\dot S_1  + 2HS_1  - S_1^2 } \right) - 2bS_2^2  - 2f_0
\left( {2A_{1} + A_{2} } \right) \nonumber \\ - 4f\left( {A_{1}^2
- A_{2}^2 } \right) - 2q_2 \left( {A_{3}^2  - A_{4}^2 } \right) =
p,
\end{eqnarray}
\begin{eqnarray}\label{5}
f\left[ {\dot A_{1}  + 2H\left( {A_{1}  - A_{2} } \right) + 4S_1
A_{2} } \right] + q_2 S_2 A_{3} \nonumber \\ - q_1 S_2 A_{4}  +
\left( {f_0  + \frac{a} {8}} \right)S_1  = 0,
\end{eqnarray}
\begin{eqnarray}\label{6}
q_2 \left[ {\dot A_{4}  + 2H\left( {A_{4}  - A_{3} } \right) + 4S_1 A_{3} } \right] - 4f\,S_2 A_{2} \nonumber \\
- 2q_1 S_2 A_{1}  - \left( {f_0  - b} \right)S_2  = 0,
\end{eqnarray}
where
\begin{eqnarray}
  a = 2a_1  + a_2  + 3a_3, \qquad b = a_2  - a_1,
\nonumber\\
  f = f_1  + \frac{{f_2 }} {2} + f_3  + f_4  + f_5  + 3f_{6}\, ,
\nonumber\\
  q_1  = f_2  - 2f_3  + f_4  + f_5  + 6f_{6}, \qquad q_2  = 2f_1  - f_2 .
\nonumber
\end{eqnarray}
The system of gravitational equations (3)-(6) allows to obtain
equations for the torsion functions $S_1$ and $S_2$, and also the
generalization of Friedmann cosmological equations for HIM in the
frame of PGTG in the case of different particular gravitational
Lagrangians. By analyzing the system of equations (3)-(6) we will
use the Bianchi identities in the Riemann-Cartan continuum, which
are reduced in the case of HIM to two following relations
\cite{2}:
\begin{equation}\label{7}
    \dot A_{2}  + 2H\left( {A_{2}  - A_{1} } \right) + 4S_1 A_{1}
        + 2S_2 A_{4}  = 0,
    \end{equation}
\begin{equation}\label{8}
\dot A_{3}  + 2H\left( {A_{3}  - A_{4} } \right) + 4S_1 A_{4}
        - 2S_2 A_{1}  = 0.
\end{equation}

In the case of gravitational Lagrangian used in \cite{1} the
following restrictions on indefinite parameters in equations
(3)-(6) are valid: $q_2=0$ and $2f=q_1$. As result by using (8) we
obtain from (6):
\begin{equation}\label{9}
\left[2f F+3(f_0-b)\right]S_2=0,
\end{equation}
where $F=6(A_1+A_2)$ is the scalar curvature. According to (9)
there are two possibilities corresponding to two different
cosmological solutions:
\begin{equation}\label{10}
S_2\neq 0 \qquad\mathrm{and}\qquad F=-\frac{3(f_0-b)}{2f},
\end{equation}
\begin{equation}\label{11}
S_2= 0 \qquad\mathrm{and}\qquad F\neq -\frac{3(f_0-b)}{2f}.
\end{equation}

In the first case, when conditions (10) are valid, the system of
equations (3)-(5) by using (7) and the definitions (2) for the
curvature functions $A_2$ and $A_1$ leads to the following
solution: $S_1=0$ and
\begin{equation}\label{12}
S_2^2  = \frac{{f_0(f_0  - b)}} {{4fb}} + \frac{{\rho  - 3p}}
{{12b }},
\end{equation}
\begin{equation}\label{13}
    \frac{k} {{R^2 }} + H^2  = \frac{1} {{6b }}\left[ {\rho  + \frac{{3\,\left( {f_0  - b} \right)^2}}
         {{4f}}} \right],
\end{equation}
\begin{equation}\label{14}
    \dot H + H^2  =  - \frac{1} {{12b }}\left[ {\rho  + 3p - \frac{{3\left( {f_0  - b} \right)^2 }}
        {{2f}}} \right].
\end{equation}
The expression (12) for the torsion function $S_2$ and
cosmological equations (13)-(14) are identical to that at
asymptotics obtained in \cite{2}. These equations allow to explain
observable acceleration of cosmological expansion by certain
restrictions on indefinite parameters \cite{2} (see also
\cite{3}); however, because the equations (13)-(14) have the form
of Friedmann cosmological equations with cosmological constant,
the problem of cosmological singularity remains in such theory.

The second possibility, when conditions (11) are valid, is
considered in \cite{1}. In this case gravitational equation (6)
vanishes, and the system of gravitational equations is reduced to
three equations (3)-(5) including two indefinite parameters: $a$
and $f$. Note that these equations take place always independently
on restrictions for indefinite parameters of gravitational
Lagrangian (1), if $S_2=0$. At first time these gravitational
equations for HIM were deduced and investigated in \cite{4}. The
analysis of gravitational equations (3)-(5) by using the Bianchi
identity (7) (with $S_2=0$) gives the following expressions for
the curvature functions $A_2$, $A_1$ and the torsion function
$S_1$ \cite{4}:
\begin{equation}\label{15}
\left.
    \begin{matrix}
    \displaystyle
        F=\frac{1}{2} \left(f_0+\frac{1}{8}a\right)^{-1}
            \left[
                \rho-3p+\frac{3}{2} a \left(\frac{k}{R^2}+\dot{H}+2H^2\right)
            \right],\\
        A_2=\frac{1}{6} \left(f_0+\frac{1}{8}a\right)^{-1}
            \frac{\rho+\frac{1}{3}f F^2+\frac{3}{4}a\left(k+\dot{R}^2\right) R^{-2}}%
                {1+\frac{2}{3}f F \left(f_0+\frac{1}{8}a\right)^{-1}},\\
        S_1=-\frac{1}{6} \left(f_0+\frac{1}{8}a\right)^{-1}
            \frac{f \dot{F}}%
                {1+\frac{2}{3}f F \left(f_0+\frac{1}{8}a\right)^{-1}},\\
        A_1=\frac{1}{6}F-A_2
            \qquad \left(f_0+\frac{1}{8}a\neq 0\right).
    \end{matrix}
\right\}
\end{equation}
Unlike GR, where gravitational equations for HIM are identical to
Friedmann cosmological equations, discussed gravitational
equations of PGTG for HIM include besides the scale factor $R$
also the torsion function $S_1$. However, gravitational equations
allow to deduce cosmological equations for the scale factor $R$
without torsion function generalizing Friedmann cosmological
equations. We obtain these equations by substituting the solution
(15) into the definitions (2) of the curvature functions $A_2$ and
$A_1$. The explicit form of the first cosmological equation is the
following:
\begin{eqnarray}
\frac{k}{R^2}+H^2  =
  \frac{1}{6f_0}\left(1+\frac{2f}{3f_0}F\right)^{-1}
  \left[
    \rho+\frac{1}{3}f F^2 -4fH \dot{F}
    \right.\nonumber\\
    \left.
   - \frac{2f^2}{3\left(f_0+a/8\right)}
        \left(1+\frac{2f}{3\left(f_0+a/8\right)}F\right)^{-1}\dot{F}^2
    \right],
    \nonumber\\
\end{eqnarray}
where the scalar curvature $F$ is defined by (15). We do not write
the second cosmological equation following from (16) by using the
conservation law, which in the case of spinless matter minimally
coupled with gravitation takes the same form as in GR:
\begin{equation}\label{17}
\dot \rho  + 3H\left( {\rho  + p} \right) = 0.
\end{equation}
Cosmological solutions for metric quantities can be found by
solving the system of equations (16)-(17) by given equation of
state for gravitating matter and initial conditions for ($R$, $H$,
$\dot H$, $\rho$). At first note that if the parameter $a$ is not
equal to zero (in \cite{1} $a=4a_1'>0$ [11]), cosmological
equations contain higher derivatives of the scale factor $R$; in
particular, the cosmological equation (16) contains the third
derivative of $R$ (the second cosmological equation is
differential equation of the fourth order with respect to $R$)
\footnote [12]{Note that gravitational equations of PGTG are
differential equations of the second order with respect to
gravitational gauge field variables -- the tetrad and the Lorentz
connection. By using these variables gravitational equations
(3)-(6) for HIM were obtained. Cosmological solutions can be found
by direct integration of gravitational equations without deduction
and consideration of cosmological equations for the scale factor
$R$ generalizing Friedmann cosmological equations of GR.
Cosmological equations are important for comparison of cosmology
built in the frame of PGTG with Friedmann cosmology of GR. As our
analysis shows, the character of cosmological solutions depends
essentially on the order of cosmological equations as differential
equations for the scale factor (see below).}. By given initial
conditions we obtain two different solutions corresponding to two
different initial values of $\ddot H$ (or $S_1$). It is because
the equation (16) contains the squared derivative $\ddot H$ and
hence by given initial values of ($R$, $H$, $\dot H$, $\rho$)
there is two different values of $\ddot H$. Cosmological equations
contain the parameter $\frac{1}{3}\frac{f}{\left(f_0
+\frac{1}{8}a\right)^{2}}\sim\
\alpha=\frac{1}{3}\frac{f}{f_0^2}>0$ with inverse dimension of
energy density, and they transform into Friedmann cosmological
equations of GR if $f\to 0$ ($\alpha\to 0$) \cite{4}. The
solutions asymptotics of cosmological equations, where energy
density is small $\alpha\rho\ll 1$, is close to the Friedmannien
asymptotics . The behaviour of cosmological solutions at present
epoch depends essentially on the value of parameter $\alpha$. If
the value of $\alpha^{-1}$ corresponds to the scale of extremely
high energy densities and hence at present epoch we have
$\alpha\rho\ll 1$, the behaviour of cosmological solutions is
quasi-Friedmannien and such solutions can not describe
cosmological acceleration. If the value of $\alpha^{-1}$ is
comparable with the average energy density at present epoch
($\alpha\rho\sim 1$), the solutions behaviour differs from that of
GR and depends essentially on values of indefinite parameters
($f$, $a$) and using initial conditions. With the purpose to
compare cosmological solutions for flat model ($k=0$) with that of
standard $\Lambda CDM$-model of GR, we will use the initial values
for ($H$, $\dot H$, $\rho$) obtained in the frame of GR and
corresponding to observational data. Note that as result by given
parameters $f$ and $a$ the initial values for $F$ and $S_1$ are
determined and they can not be introduced independently from $f$
and $a$ (compare with \cite{1}).

Now let us to demonstrate these statements by numerical solution
of Eqs. (16)-(17) in the case of flat model filled with the dust
matter ($p=0$), for which $\rho R^3=\mathrm{const}$. With this
purpose dimensionless units for variables and parameters will be
introduced by the following way: $t\to
\tilde{t}=t\sqrt{\frac{\rho_{\mathrm{cr}}}{6f_0}}$,
$H\to\tilde{H}=H\sqrt{\frac{6f_0}{\rho_{\mathrm{cr}}}}$,
$\dot{H}\to \tilde{H}'=\frac{d\tilde{H}}{d\tilde{t}}$,
$\rho\to\tilde{\rho}=\frac{\rho}{\rho_{\mathrm{cr}}}$,
$a\to\tilde{a}=\frac{a}{f_0}$ and
$f\to\tilde{f}=f\frac{\rho_{\mathrm{cr}}}{3f_0^2}$, where the
value of $\rho_{\mathrm{cr}} = 6f_0 H_0^2$ corresponds to average
energy density in the Universe at present epoch in the frame of GR
(the index "$0$" at physical quantities denotes its values at
present epoch). In the frame of standard $\Lambda CDM$-model one
supposes that the total value of energy density includes the
contributions of three components: baryonic matter, dark matter
and dark energy. By taking into account that one uses at present
epoch for baryonic and dark matter the equation of state of dust
($p_{\mathrm{BM}}=p_{\mathrm{DM}}=0$), and for dark energy
$p_{\mathrm{DE}}=-\rho_{\mathrm{DE}}$, we write the Friedmann
cosmological equations of GR in dimensionless form for considering
case:
\begin{eqnarray}\label{18}
&& \tilde{H}^2  = \tilde{\rho}_{\mathrm{BM}} +
\tilde{\rho}_{\mathrm{DM}} +
\tilde{\rho}_{\mathrm{DE}},\nonumber \\
&& \tilde{H}' + \tilde{H}^2  =  - \frac{1}{2}
(\tilde{\rho}_{\mathrm{BM}} +
\tilde{\rho}_{\mathrm{DM}}-2\tilde{\rho}_{\mathrm{DE}}),
\end{eqnarray}
In accordance with Eqs. (18) and observational data we have the
following initial values for physical parameters: $\tilde{H}_0=1$,
$\tilde{\rho}_{\mathrm{BM0}} + \tilde{\rho}_{\mathrm{DM0}}=0.3$,
$\tilde{H'}_0=-0.45$. If one supposes that there is in the
Universe only baryonic and dark matter ($\rho=\rho_{\mathrm{BM}} +
\rho_{\mathrm{DM}}$), then we obtain the following initial
condition for energy density $\tilde{\rho}_0=0.3$.

By using obtained initial conditions numerical solutions for
dimensionless Hubble parameter and acceleration $w=\tilde{H}' +
\tilde{H}^2$ are presented in Fig. 1 - 2 for the following choose
of parameters: $\tilde{f}=0.35$ and $\tilde{a}=2$.

\begin{figure}
 \includegraphics{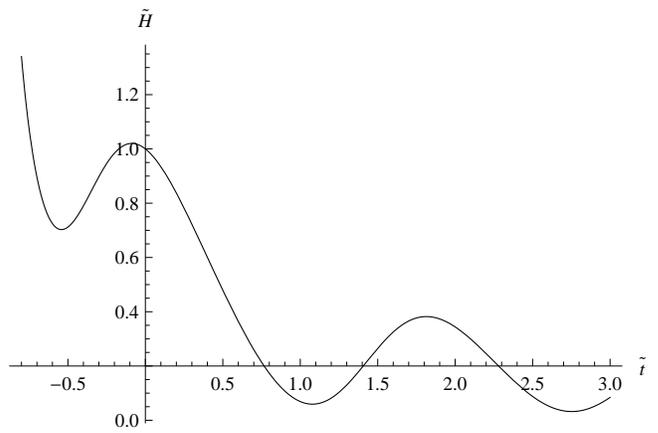}
 \caption[]{Solution for the Hubble parameter by initial
 conditions: $\tilde{\rho}_0=0.3$,
 $\tilde{H}_0=1$, $\tilde{H}'_0=-0.45$, $\tilde{H}''_0= -4{.}60$ and $\tilde{a}=2$, $\tilde{f}=0.35$.}
\end{figure}

\begin{figure}
 \includegraphics{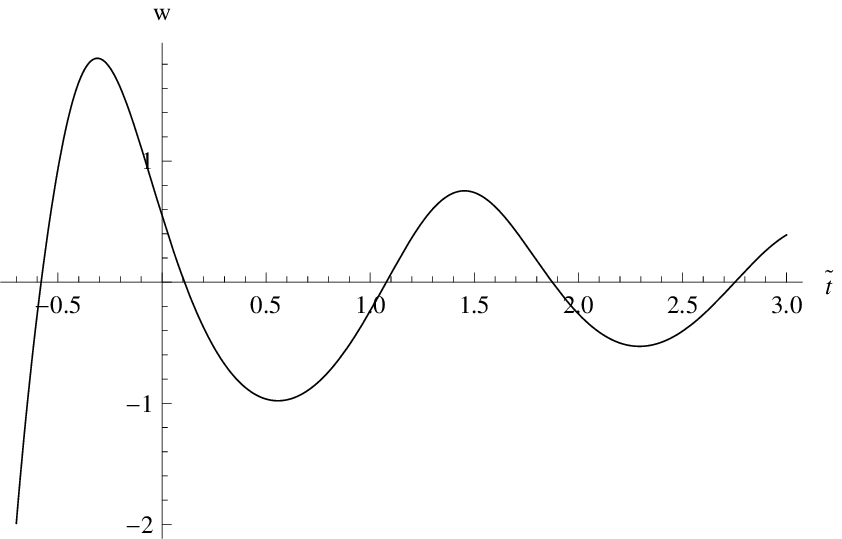}
 \caption[]{Solution for acceleration $w$ by initial
 conditions: $\tilde{\rho_0}=0.3$,
 $\tilde{H}_0=1$, $\tilde{H_0'}=-0.45$, $\tilde{H}''_0= -4{.}60$ and $\tilde{a}=2$, $\tilde{f}=0.35$.}
\end{figure}

\begin{figure}
 \includegraphics{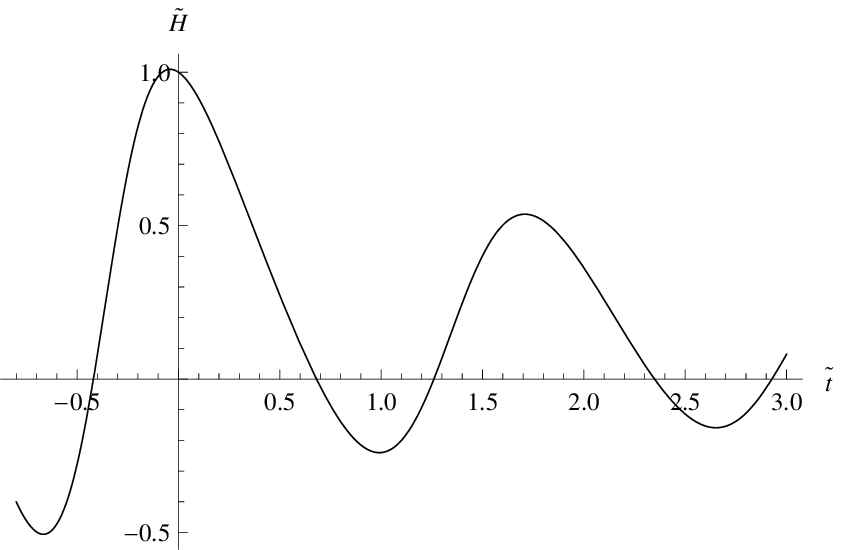}
 \caption[]{Solution for the Hubble parameter by initial
 conditions: $\tilde{\rho}_0=0.05$,
 $\tilde{H}_0=1$, $\tilde{H}'_0=-0.45$, $\tilde{H}''_0= -7{.}97$ and $\tilde{a}=2$, $\tilde{f}=0.35$.}
\end{figure}

\begin{figure}
 \includegraphics{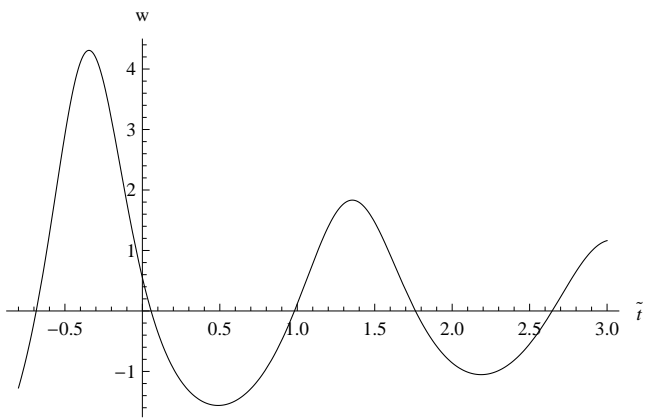}
 \caption[]{Solution for acceleration $w$  by initial
 conditions: $\tilde{\rho}_0=0.05$,
 $\tilde{H}_0=1$, $\tilde{H}_0'=-0.45$, $\tilde{H}''_0= -7{.}97$ and $\tilde{a}=2$, $\tilde{f}=0.35$.}
\end{figure}

Similar solutions were used in \cite{1}. Note that the solution
presented in Fig. 1-2 corresponds to the following initial value
of $\tilde{H}''$ (or $\tilde{S}_1$): $\tilde{H}''_0= -4{.}60$
($\tilde{S}_{10}= 0{.}185$). By using the second initial value of
$\tilde{H}''_0= -32{.}5$ ($\tilde{S}_{10}= 0{.}815$) compatible
with used initial values for $\tilde{H}$, $\tilde{H'}$ and
$\tilde{\rho}$, we obtain not physical solution with non-
Friedmannien asymptotics -- with rapidly growing values of energy
density and the Hubble parameter in the future. Although by virtue
of used initial conditions the solutions similar to that presented
in Fig. 1-2 have at present epoch the same observational
characteristics as solutions of the standard $\Lambda CDM$-model,
however, the behaviour of cosmological solutions in the past as
well as in the future differs essentially from that of GR. Unlike
$\Lambda CDM$-model, where the transition from decelerating to
accelerating expansion ($w=0$) takes place at certain moment in
the past, in the case of discussed solutions the acceleration
function $w$ oscillates near $w=0$. The character of solutions of
eq. (16) with quasi-Friedmannien asymptotics in the future depends
essentially on initial values for $\tilde{H}$, $\tilde{H'}$ and
$\tilde{\rho}$. As illustration of this statement the solution of
(16)-(17) corresponding to the following initial conditions:
$\tilde{\rho}_0=0.05$, $\tilde{H}_0=1$, $\tilde{H}'_0=-0.45$,
$\tilde{H}''_0= -7{.}97$ ($\tilde{S}_{10}= 0{.}261$) is presented
in Fig. 3-4. Such initial conditions correspond to supposition
that at present all matter in the Universe is reduced to the
baryonic matter. We see that according to Fig. 3 the evolution in
this case has the oscillating character. The second solution
corresponding to used initial values of $\tilde{\rho}$,
$\tilde{H}$, $\tilde{H}'$ is also not physical solution with non-
Friedmannien asymptotics.

We see that in the case $a\neq 0$ the character of cosmological
evolution differs essentially from that of GR. This is connected
with the presence in cosmological equations of higher derivatives
of $R$. In order to exclude higher derivatives from cosmological
equations the condition $a=0$ was used in Refs. \cite{4} and
\cite{2}. Then according to (15) the torsion $S_1$ is certain
function of energy density and pressure, and all cosmological
solutions of generalized cosmological Friedmann equations obtained
in \cite{4} are physical solutions with usual Friedmannien
asymptotics. These equations allow to build regular isotropic
cosmology including inflationary cosmology \cite{5,6,7}. Regular
character of all cosmological solutions with respect to metrics,
Hubble parameter, its time derivative and energy density is
connected with gravitational repulsion effect provoked by the
torsion function $S_1$ at extreme conditions in the beginning of
cosmological expansion. However, the torsion function $S_1$ tends
to zero at asymptotics, and such theory can not describe the
accelerating Universe. As was shown in \cite{2}, regular HIM with
quasi-Friedmannien accelerating cosmological expansion at present
epoch contain two torsion functions $S_1$ and $S_2$, and the
effect of gravitational repulsion at present epoch is connected
with effective cosmological constant induced by the torsion
function $S_2$. As it follows from our consideration given above,
the gravitational Lagrangian adopted in \cite{1} allows to build
HIM only with one torsion function. In the case of HIM with
$S_2=0$ the scalar curvature $F$ is not constant for physical
solutions (see (15)). However, if the dominator in (15) is equal
to zero (that corresponds to constant scalar curvature),
gravitational equations do not allow to determine the torsion
function $S_1$ and the curvature functions (compare with
\cite{1}). Gravitational equations have in this case the following
specific solution: cosmological equations take the form of
Friedmann cosmological equations with cosmological constant, in
which instead of Newton's gravitational constant figures the
parameter inversely proportional to $(-a)$. If $a>0$, this
specific solution is not physical. In connection with this the
sign of $a$ for this case was changed that leads to "degenerate"
case discussed in \cite{1}. Note that specific not physical
solution with constant scalar curvature does not appear if $a=0$.

The investigation of isotropic cosmology in the frame of PGTG
based on gravitational Lagrangian (1) shows that the PGTG can have
principal meaning for theory of gravitational interaction \cite{8,
9}. Physical consequences depend essentially on restrictions on
indefinite parameters of $\mathcal{L}_{\rm g}$. Because at present
the quadratic part of $\mathcal{L}_{\rm g}$ is unknown, we have to
investigate the PGTG by using general expression (1) of
gravitational Lagrangian and to obtain restrictions on indefinite
parameters of $\mathcal{L}_{\rm g}$ in order that physical
consequences and mathematical structure of gravitational equations
should be satisfactory.

As it is follows from our consideration given above the PGTG based
on gravitational Lagrangian used in \cite{1} (and also in
\cite{10}) lead to isotropic cosmology, in the frame of which the
evolution of the Universe differs essentially from that of
standard $\Lambda CDM$-model. Unlike \cite{1} cosmological
equations deduced in \cite{2} take at asymptotics the
quasi-Friedmannien form, and terms related to dark energy (and
possibly to dark matter \cite{3}) in cosmological equations of
general relativity theory for $\Lambda CDM$-model are connected in
considered theory with the change of gravitational interaction
provoked by spacetime torsion.

One of the authors (A.V.M.) is very grateful to Professor F.W.
Hehl for discussions which initiated this Comment.


\end{document}